\documentclass{aa}
\usepackage{txfonts}
\usepackage{natbib}
\usepackage{graphicx}
\usepackage[below]{placeins}

\bibpunct{(}{)}{;}{a}{}{,}

\begin{document}

\title{High-resolution X-ray spectroscopy of the Narrow line Seyfert 1
galaxy NGC~4051 with Chandra LETGS}

\author{M. Fe\v{n}ov\v{c}\'\i{}k\inst{1,2}
 \and J.S. Kaastra\inst{1}
 \and E.~Costantini\inst{1,2}
 \and K.C.~Steenbrugge\inst{3}
 \and F. Verbunt\inst{2}}

\offprints{J.S. Kaastra}

\institute{     SRON Netherlands Institute for Space Research,
Sorbonnelaan 2,
                NL - 3584 CA Utrecht, the Netherlands
         \and
                Astronomical Institute, Utrecht University, P.O. Box
80000,
                NL - 3508 TA Utrecht, the Netherlands
         \and
	 St John's College Research Centre fellow, University of Oxford,
	 Oxford, OX1 3JP, UK 
	 }

\date{\today}

\abstract
{ With the new generation of high-resolution X-ray spectrometers the
understanding of warm absorbers in Active Galactic Nuclei has improved
considerably. However, the main question remains the distance and structure
of the photoionised wind. }
{ We study the absorption and emission properties of the photoionised gas
near one of the brightest and most variable AGN, the Seyfert galaxy NGC~4051, in
order to constrain the geometry, dynamics and ionisation structure of the
outflow.} 
{ We analyse two observations taken with the Low Energy Transmission Grating
Spectrometer (LETGS) of Chandra. We study the spectra of both observations
and investigate the spectral response to a sudden, long-lasting flux decrease
of a factor of 5 that occurred during the second observation. }
{ We confirm the preliminary detection of a highly ionised component with an
outflow velocity of $-4500$~km\,s$^{-1}$, one of the highest velocity outflow
components seen in a Seyfert 1 galaxy. The sudden drop in intensity by a factor
of five during the second observation causes a drop in ionisation parameter of a
similar magnitude in the strongest and main ionisation component ($v =
-610$~km\,s$^{-1}$), allowing us for the first time to determine the
recombination time of this component and thereby its distance in a robust way.
We find an upper limit to the distance of $10^{15}$~m, ruling out an origin in
the narrow emission line region. In addition, an emission component producing
strong radiative recombination continua of \ion{C}{vi} and \ion{C}{v} appears
during the low state. This can be explained by emission from an  ionised skin of
the accretion disk at a distance of only $\sim 4\times 10^{12}$~m from the black
hole. Finally, the spectra contain a broad relativistic \ion{O}{viii} line with
properties similar to what was found before in this source with XMM-Newton; this
line has disappeared during the low flux state,  consistent with the
disappearance of the inner part of the accretion disk during that low flux
state. } 
{ Combining high-resolution spectroscopy with timing information, we
have constrained the geometry of the emission and absorption components
in NGC~4051.}


\titlerunning{High-resolution X-ray spectroscopy of NGC~4051}
\authorrunning{M. Fe\v{n}ov\v{c}\'\i{}k et al.}

\maketitle

\begin{figure*}[t]
\begin{minipage}{0.5\textwidth}
\includegraphics[width=0.7\textwidth,clip=t,angle=270.]{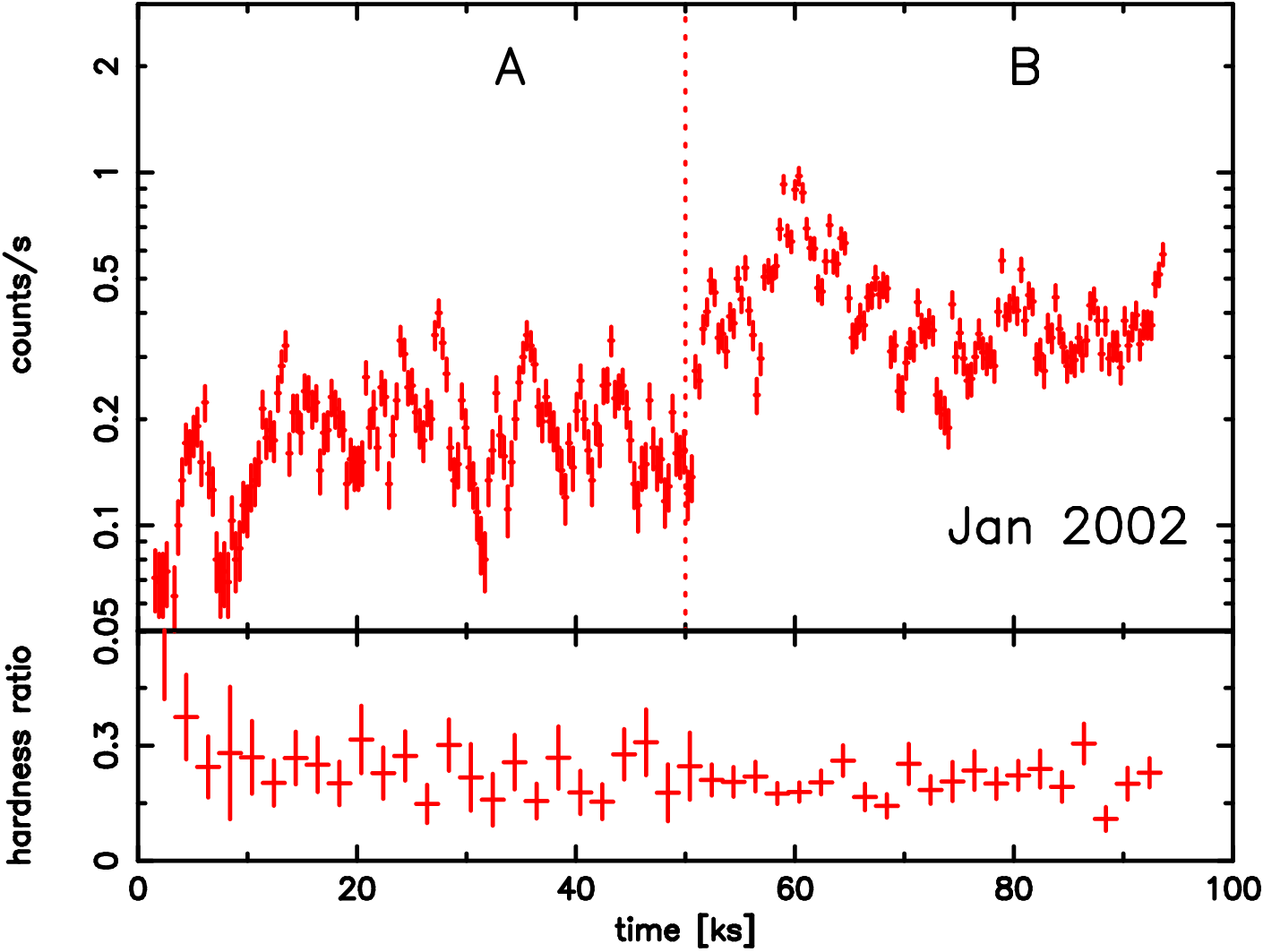}
\end{minipage}
\begin{minipage}{0.5\textwidth}
\includegraphics[width=0.7\textwidth,clip=t,angle=270.]{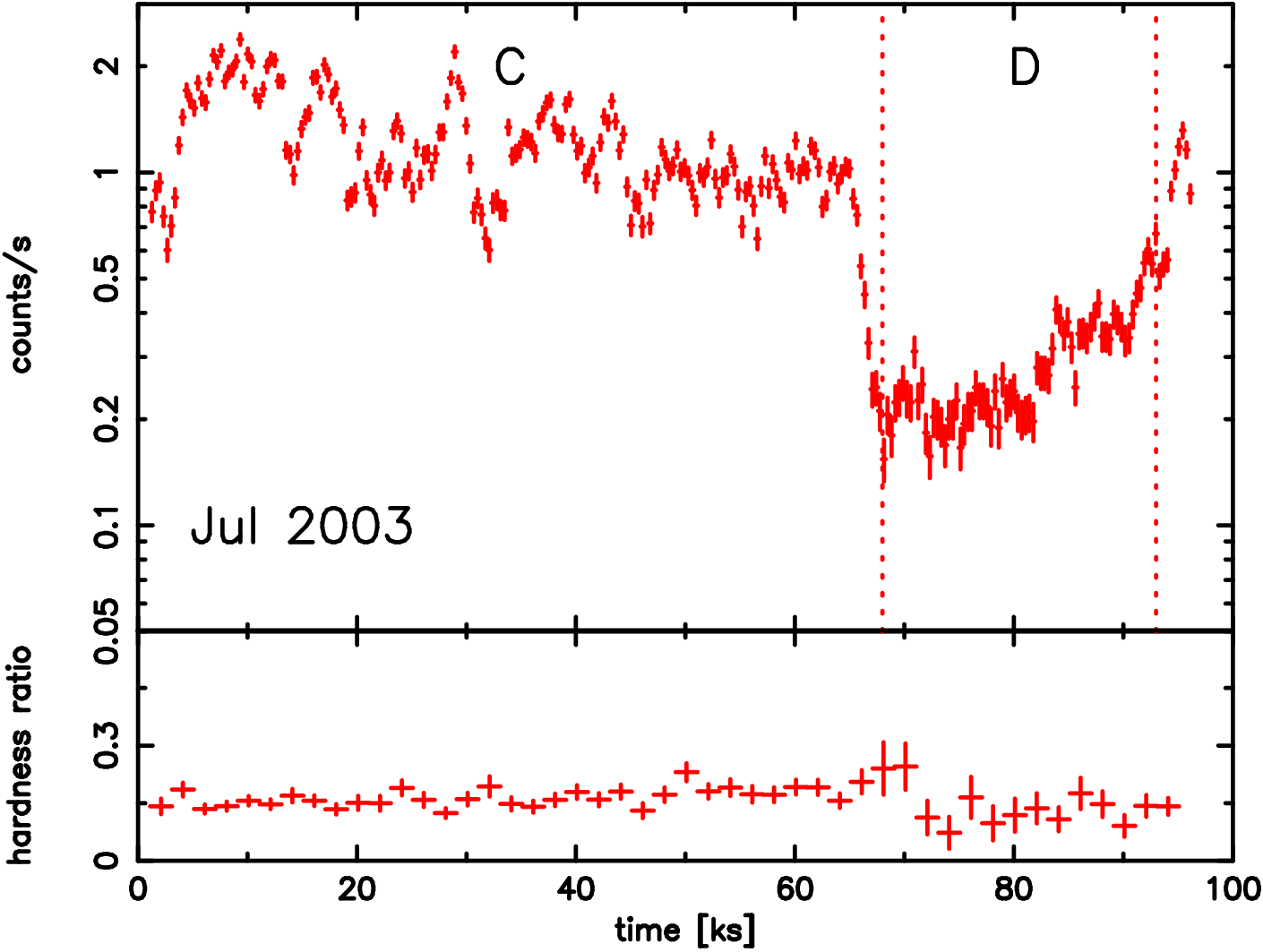}
\end{minipage}
\caption{The light curve in the zeroth spectral order (top panels)
and hardness ratio (bottom panels) for the observations of
Jan 2002 and July 2003. The
hardness ratio is defined as the ratio of the 2 -- 8 and 8 -- 32 \AA{}
counts. The top panel has a bin size of 350~s and the bottom panel has
a bin size of 2000~s. Dotted lines denote the different flux states
during the observations.}
\label{fig.lightcurves}
\end{figure*}

\section{Introduction}

The stupendous amount of energy emitted by an Active Galactic Nucleus is
released by gas that flows towards a supermassive black hole in the centre of a
galaxy, presumably via an accretion disk. The average  energy of the emitted
radiation increases towards the centre of the accretion disk, and thus X-rays
provide the best probe of the gas flow in the immediate surroundings of the
black hole. From the presence of jets, as well as from blue-shifted ultraviolet
absorption lines, we learn that in addition to the flow towards the black hole,
there is also gas flowing away from it. With the availability of X-ray
spectrographs on Chandra and XMM-Newton, we can study these flows close to the
black hole, by studying  lines in the X-ray spectra. The presence of multiple
absorption line systems, that differ in their level of ionisation or in their
outflow velocity, constrains the geometry of the outflow (for example NGC~3783,
\citet{netzer}; NGC~5548, \citet{steenbrugge}). 

In this paper we describe the X-ray spectra of the Active Galactic Nucleus of
NGC\,4051, obtained with the Low-Energy Transmission Grating on board of the
Chandra satellite. Emission from the nucleus of NGC\,4051 was already found by
\citet{hubble}, and the galaxy is one of the 6 observed by \citet{seyfert}, and
discussed in his seminal paper as one of the first 12 known Seyfert galaxies.
Being a member of the Ursa Major cluster \citep{tullyetal} it has a distance of
18.6~Mpc \citep{tullypierce} and a redshift of 700~km\,s$^{-1}$
\citep{verheijen}. NGC\,4051 is a relatively bright optical object, at
$V\simeq13.5$. In X-rays, the source was not detected with UHURU or Ariel-V, but
was studied extensively with EXOSAT and ROSAT; its X-ray variability was
recently analysed on the basis of RXTE and XMM-Newton data \citep{mchardy}. Its
X-ray continuum has been described with a power law with photon index
$\Gamma=$1.8-2.0 \citep{nandrapounds}. A soft excess with respect to this power
law is not well described by a multi-temperature disk, and \citet{ogle}
concluded that it is due to broad emission from relativistic \ion{O}{vii}
emission lines and to radiative recombination.

Two absorption line systems, at $-2340\pm 130$ and $-600\pm130$~km\,s$^{-1}$,
were found in the X-ray spectrum with the High Energy Transmission Grating of
Chandra \citep{collinge}. Multiple absorption systems were also found in the
ultraviolet with the Space Telescope Imaging Spectrograph and in the Far
Ultraviolet with the Far Ultraviolet Spectroscopic Explorer; one of these may
coincide with the $-600\pm130\,\mathrm{km\,s}^{-1}$ X-ray absorption system, all
the others have smaller absolute velocities \citep{collinge,kaspi}.

Interestingly, a preliminary analysis of our first LETGS spectrum
\citet{vandermeer} did not show that high velocity component, but indicated the
presence of an outflow component at even higher velocity, $-4500$~km\,s$^{-1}$.
For that reason we requested our second observation. This second observation,
taken when NGC\,4051 was in a much higher flux state, allowed us to investigate
the ionised outflow both in terms of outflow velocity and ionisation structure.
In this paper we report the analysis of both observations with the LETGS. 

\section{Observations and data reduction}

In this paper we describe two observations of NGC\,4051 with the Low Energy
Transmission Grating Spectrometer (LETGS, \citet{brinkman}) of Chandra. The
detector used was the High-Resolution Camera (HRC-S). The observations were done
from December 31, 2001 17:40:41 to January 1, 2002 19:51:46 and July 23, 2003
00:34:15 to July 24 03:24:43, and with 94.2 and 96.6\,ks net exposure time,
respectively. Below we refer to these observations as the Jan 2002 and Jul 2003
observation, respectively. The data reduction is described in detail by
\citet{kaastra1}. Basically, after event selection we produce background
subtracted spectra and light curves. As the HRC-S detector has limited energy
resolution, higher spectral orders cannot be separated in the data but are taken
into account properly during spectral fitting.

\section{Data analysis and results}

For all spectral fitting we used SPEX code developed at SRON \citep{kmn}.
Details of the spectral model and the time intervals for which spectral fits are
made are given later in this paper. We find that each spectrum requires several
components to be fitted adequately. To minimise the number of components, we add
them one by one, until the added component does not improve the fit
significantly. It will be noted that the continuum emission components and the
absorption components describe the global spectrum, whereas the narrow and
broadened line emission features are limited to specific wavelengths. We
therefore determined errors on the continuum emission and absorption systems
while keeping the emission line parameters fixed to the values of the best fit;
and determined errors on the emission line features while keeping the continuum
and absorption lines fixed to the values of the best fit. The fit parameters are
listed in Table\,\ref{table01}. All models were corrected for Galactic
absorption with a column density of $1.31\times 10^{24}$~m$^{-2}$ \citep{elvis}
and also for the small cosmological redshift of 700~km\,s$^{-1}$. Finally,
Table~\ref{deltachi} gives the increase in $\chi^2$ when omitting a given
spectral component from the final model. All errors correspond to $\Delta\chi^2
= 2$ (84~\% confidence).

\begin{table*}
\caption{The parameters of the spectral components for the January 2002 whole
observation (plus A and B part) and July 2003 whole observation (plus C and D
part). The time intervals A to D are defined in Fig. \ref{fig.lightcurves}. 
All wavelengths and velocities refer to the rest frame of NGC~4051.}
 \begin{tabular}{lcccccc}
\hline
\hline
						     &January 2002	     &A part		       &B part  	       &July 2003	       &C part  	       &D part\\					       
\hline
{\bfseries{Continuum}}  			     &  		     &  		       &		       &		       &		       &\\
{$\bullet\ $\textit{power--law}}		     &  		     &  		       &		       &		       &		       &\\
\ \ \ \ Flux (2--10 keV)$^1$			     &$0.96\pm 0.04$	     &$0.83\pm 0.12$	       &$1.19\pm 0.05$         &$1.92\pm 0.07$         &$2.57\pm0.06$	       &$0.41\pm 0.03$\\
\ \ \ \ Photon index $\Gamma$			     &$1.85\pm{0.04}$	     &$1.58^{+0.11}_{-0.21}$   &$2.01\pm{0.05}$        &$2.24\pm{0.03}$        &$2.23\pm{0.02}$        &$2.22\pm{0.13}$\\
{$\bullet\ $\textit{modified black--body}}           &  		     &  		       &		       &		       &		       &\\
\ \ \ \ Flux$^2$				     &$1.2\pm{0.2}$	     &$1.0\pm{0.3}$	       &$1.2\pm{0.2}$	       &$2.2\pm{0.3}$	       &$2.8\pm{0.3}$	       &$1.0\pm{0.2}$\\
\ \ \ \ kT (keV) 				     &$0.102\pm{0.003}$      &$0.103^{+0.017}_{-0.007}$&$0.103\pm{0.005}$      &$0.143\pm{0.005}$      &$0.149\pm{0.003}$      &$0.179\pm{0.020}$\\
\hline
{\bfseries{Absorption components}}		     &  		     &  		       &		       &		       &		       &\\
{$\bullet\ $\textit{warm absorber 1$^3$}}	     &  		     &  		       &		       &		       &		       &\\	      
\ \ \ \ N$_{\rm H}$				     &---		     &---		       &---		       &$0.34\pm{0.05}$        &$0.41\pm{0.04}$        &---\\
\ \ \ \ $\log\xi$				     &---		     &---		       &---		       &$0.67\pm{0.06}$        &$0.74\pm{0.06}$        &---\\
\ \ \ \ v$_{turb}$				     &---		     &---		       &---		       &$150^{+40}_{-20}$      &$150\pm{20}$	       &---\\
\ \ \ \ v$_{out}$				     &---		     &---		       &---		       &$-210\pm{20}$	       &$-210\pm{20}$	       &---\\
{$\bullet\ $\textit{warm absorber 2$^3$}}	     &  		     &  		       &		       &		       &		       &\\	      
\ \ \ \ N$_{\rm H}$				     &$0.7\pm{0.3}$          &$1.95^{+22.0}_{-1.39}$   &$0.6\pm{0.2}$          &$0.6\pm{0.3}$          &$1.9\pm{1.0}$	       &$3.5\pm{1.0}$\\
\ \ \ \ $\log\xi$				     &$1.80\pm{0.25}$	     &$2.37\pm{0.26}$	       &$1.40\pm{0.18}$        &$2.24\pm{0.10}$        &$2.44\pm{0.08}$        &$1.54\pm{0.10}$\\
\ \ \ \ v$_{turb}$				     &$75\pm{30}$	     &$70^{+80}_{-50}$         &$90\pm{40}$            &$90^{+40}_{-20}$       &$40\pm{25}$	       &$50\pm{20}$\\
\ \ \ \ v$_{out}$				     &$-620\pm{60}$	     &$-680^{-630}_{+190}$     &$-570\pm{80}$	       &$-570\pm{60}$	       &$-610\pm{40}$	       &$-580\pm{40}$\\
{$\bullet\ $\textit{warm absorber 3$^3$}}	     &  		     &  		       &		       &		       &		       &\\		
\ \ \ \ N$_{\rm H}$				     &$19^{+15}_{-8}$	     &$74^{+74}_{-51}$         &$14\pm{8}$	       &$20\pm{10}$	       &$15^{+10}_{-6}$        &---\\
\ \ \ \ $\log\xi$				     &$3.2\pm{0.1}$	     &$3.4^{+0.7}_{-0.1}$      &$3.0\pm{0.1}$	       &$3.1^{+0.1}_{-0.3}$    &$3.2\pm{0.1}$	       &---\\
\ \ \ \ v$_{turb}$				     &$160\pm{60}$	     &$160^{1230}_{-30}$       &$130^{+70}_{-40}$      &$11^{+100}_{-11}$      &$15\pm{15}$	       &---\\
\ \ \ \ v$_{out}$				     &$-4450\pm{110}$	     &$-4310^{+720}_{-520}$    &$-4570\pm{120}$        &$-4670^{+100}_{-1000}$ &$-4600\pm{120}$        &---\\
\hline
{\bfseries{Broad emission lines}}		     &  		     &  		       &		       &		       &		       &\\
{$\bullet\ $\textit{\ion{O}{viii} 
   Ly$\alpha$ line$^4$}}			     &  		     &  		       &		       &		       &		       &\\
\ \ \ \ flux					     &$12.7\pm{1.8}$	     &$9.7\pm{1.8}$	       &$15.9\pm{2.6}$         &$6.9\pm{0.7}$          &$7.7^{+2.4}_{-1.0}$    &$1.0\pm{0.4}$\\
\ \ \ \ $\lambda$				     &$18.7^{+5.7}_{-1.9}$   &$18.5^{+0.2}_{-1.7}$    &$19.0^{+2.2}_{-1.0}$    &$19.26^{+0.84}_{-0.10}$&$19.08^{+1.14}_{0.12}$ &$18.77\pm{0.16}$\\
\ \ \ \ FWHM					     &[21]		     &[21]		       &[21]		       &[2.5]		       &[3.2]		       &$0.69\pm{0.32}$\\
\ \ \ \ q					     &$5.2\pm{0.6}$	     &$5.2^{+1.0}_{-0.5}$      &$5.2\pm{0.6}$	       &$1.9^{+0.2}_{-0.8}$		       &$2.0^{+0.2}_{-1.2}$    &---\\
\ \ \ \ $i$					     &$48^{+1}_{-22}$	     &$49^{+13}_{-49}$         &$49^{+5}_{-29}$        &$49^{+40}_{-1}$        &$46^{+7}_{-1}$         &---\\		      
{$\bullet\ $\textit{\ion{O}{vii} triplet$^4$}}       &  		     &  		       &		       &		       &		       &\\
\ \ \ \ flux					     &---		     &---		       &---		       &$3.8\pm{0.7}$	       &$5.0^{+1.1}_{-1.8}$   &$1.4\pm{0.4}$\\
\ \ \ \ $\lambda$				     &---		     &---		       &---		       &$22.07\pm{0.14}$       &$22.07\pm{0.16}$       &$21.72\pm{0.02}$\\
\ \ \ \ FWHM					     &---		     &---		       &---		       &$1.50\pm{0.30}$        &$1.61\pm{0.40}$        &$0.16\pm{0.06}$\\
{$\bullet\ $\textit{\ion{C}{vi} Ly$\alpha$ line$^4$}}&  		     &  		       &		       &		       &		       &\\
\ \ \ \ flux					     &---		     &---		       &---		       &$2.3\pm{0.4}$	       &$2.4\pm{0.6}$	       &$1.8\pm{0.5}$\\
\ \ \ \ $\lambda$				     &---		     &---		       &---		       &$33.68\pm{0.03}$       &$33.70\pm{0.05}$       &$33.68\pm{0.04}$\\
\ \ \ \ FWHM					     &---		     &---		       &---		       &$0.30\pm{0.08}$        &$0.32\pm{0.13}$        &$0.25\pm{0.07}$\\
\hline
{\bfseries{Narrow emission features}}		     &  		     &  		       &		       &		       &		       &\\
{$\bullet\ $\textit{\ion{O}{vii} forbidden line$^4$}}&  		     &  		       &		       &		       &		       &\\
\ \ \ \ flux					     &$0.76\pm{0.12}$	     &$0.9\pm{0.2}$	       &$0.8\pm{0.2}$	       &$1.0\pm{0.2}$	       &$0.9\pm{0.2}$	       &$1.4\pm{0.2}$\\
\ \ \ \ $\lambda$				     &$22.089\pm{0.007}$     &$22.089\pm{0.006}$       &$22.087\pm{0.013}$     &$22.095\pm{0.005}$     &$22.093\pm{0.006}$     &$22.090\pm{0.012}$\\
{$\bullet\ $\textit{RRC$^5$}}			     &  		     &  		       &		       &		       &		       &\\  
\ \ \ \ T					     &$3.2^{+2.7}_{-1.0}$    &$2.9^{+6.8}_{-1.4}$      &$3.2^{+3.0}_{-1.1}$    &---		       &---		       &$4.9\pm{1.8}$\\
\ \ \ \ EM$_\mathrm{\ion{C}{v}}$		     &$0\pm{20}$             &$0\pm{100}$	       &$0\pm{20}$	       &---		       &---		       &$180\pm{100}$\\
\ \ \ \ EM$_\mathrm{\ion{C}{vi}}$		     &$40\pm{30}$            &$20^{+30}_{-20}$         &$70\pm{50}$  	       &---		       &---		       &$320\pm{120}$\\
\ \ \ \ EM$_\mathrm{\ion{N}{vi}}$		     &$10^{+20}_{-10}$	     &$10^{+40}_{-10}$	       &$0\pm{20}$	       &---		       &---		       &$30\pm{40}$\\
\ \ \ \ EM$_\mathrm{\ion{N}{vii}}$		     &$0\pm{10}$	     &$0\pm{10}$	       &$7^{+16}_{-7}$	       &---		       &---		       &$0\pm{20}$\\
\ \ \ \ EM$_\mathrm{\ion{O}{vii}}$		     &$7\pm{7}$ 	     &$7^{+11}_{-7}$	       &$11\pm{11}$	       &---		       &---		       &$16\pm{15}$\\
\ \ \ \ EM$_\mathrm{\ion{O}{viii}}$		     &$0\pm{5}$	             &$0\pm{5}$		       &$0\pm{15}$	       &---		       &---		       &$10\pm{10}$\\
\hline
\end{tabular}
\\
\noindent
$^1$ Absorption-corrected flux in $10^{-14}$ W\,m$^{-2}$\\
$^2$ Absorption-corrected bolometric flux in $10^{-14}$ W\,m$^{-2}$\\
$^3$ The column density N$_{\rm H}$ is in units of $10^{25}$ m$^{-2}$. The ionisation parameter $\xi$ is
in units of $10^{-9}$ W\,m, the turbulent velocity v$_{turb}$ and the outflow
velocity v$_{out}$ are in km\,s$^{-1}$\\
$^4$ The flux is in ph\,m$^{-2}$\,s$^{-1}$, the wavelength
and the FWHM are in \AA. In the case of the relativistic profile, $i$ is in
degrees. FWHM values in [] were not fitted but computed from the best-fit
relativistic line profile.\\
$^5$ the temperature is in eV and the emission measure is in units of
$10^{64}$ m$^{-3}$\\
    \label{table01}
\end{table*}

\begin{table*}
\begin{center}
\caption{Increase in $\chi^2$ when a given spectral component is omitted
from the model.}
 \begin{tabular}[t]{lccccccc}
   \hline
   \hline
omitted component  & number of    &Jan 2002  &A      &B        &July 2003   &C	      &D\\
                   & parameters   &          &       &         &            &         &\\    
\hline
warm absorber 1     &4 &---	  &---      &---      &$380.56$      &$413.94$      &---\\
warm absorber 2     &4 &$41.05$   &$7.06$   &$51.54$  &$102.08$      &$80.34$	    &$55.58$\\
warm absorber 3     &4 &$76.27$   &$34.12$  &$54.36$  &$59.89$       &$20.63$	    &---\\
O VIII Ly$\alpha$ line&4 or 3&$41.04$   &$25.76$ &$27.71$   &$54.34$       &$45.28$	    &$7.51$\\
O VII triplet	    &3 &---	  &---      &---      &$34.84$       &$45.28$	    &$17.23$\\
C VI Ly$\alpha$ line&3 &---	  &---      &---      &$22.79$       &$31.39$	    &$12.38$\\
O VII forbidden line&2 &$57.65$   &$36.94$  &$23.03$  &$43.74$       &$23.80$	    &$29.37$\\
RRC		     &7&$7.95$    &$3.66$   &$5.68$   &---       &---	    &$30.92$\\
\hline
$\chi^2$ of full model &&$1614.84$ &$1603.74$&$1602.77$&$2820.43$     &$2755.70$    &$2144.82$\\
d.o.f.                 && 1478     & 1478    & 1478    & 2563        & 2563  & 1591 \\  
\end{tabular}
\label{deltachi}
\end{center}
\end{table*}

\subsection{Flux level variations}

\begin{figure*}[!htb]
\begin{center}
\includegraphics[angle=-90, width=0.9\textwidth]{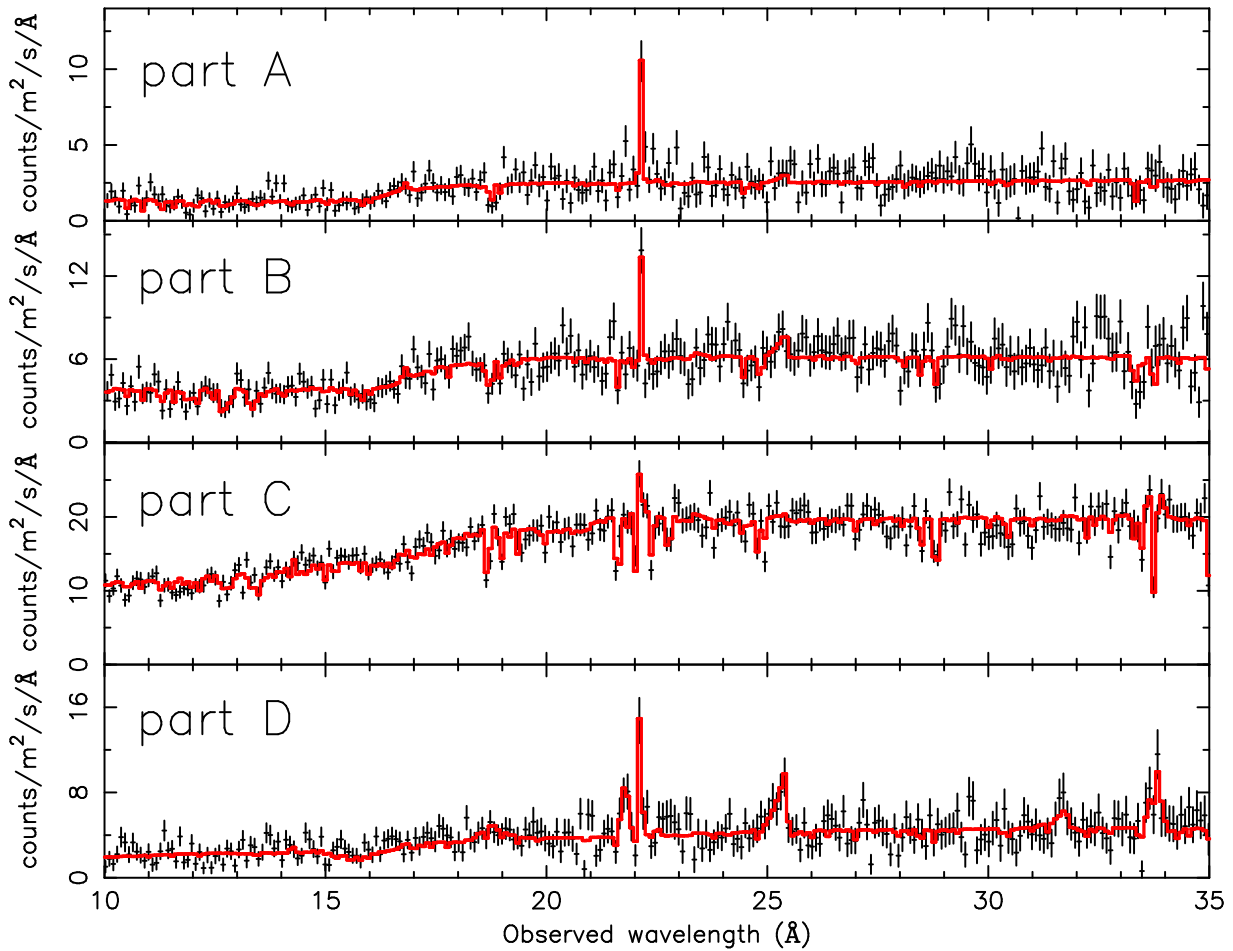}
\caption{Flux spectrum and our best spectral model for each individual light
curve part, which are marked as depicted on Fig.~\ref{fig.lightcurves}. Note the
presence of strong RRCs of \ion{C}{vi} at 25.43 \AA{} and of \ion{C}{v} at 31.80
\AA{} in spectrum D. The \ion{O}{vii} forbidden line at 22.10 \AA{} is visible
in every spectrum and has the same strength in spectra A, B and C, but it is
stronger in spectrum D. Note the different continuum level and shape in spectrum
C compared to the other spectra.}
\label{spectrum_ABCD}
\end{center}
\end{figure*}

NGC\,4051 was variable in both our observations, as shown in
Fig.~\ref{fig.lightcurves}, where we plot the count rate obtained with the LETGS
in zeroth spectral order. During the Jan 2002 observation the average brightness
of the source increased about halfway the observation to a somewhat higher
level; during the Jul 2003 observation the source was brighter yet during the
first $\sim$ 65\,000\,s and then dropped in about 3000\,s to the level of the
second half of our first observation. It then recovered gradually to the high
level again. On top of these long-term variations we also detect variability
with smaller, but significant, amplitude on shorter time scales of a few
thousand seconds. These rapid variations are smaller after the rapid flux
decrease of the Jul 2003 observation. In the analysis below we will discuss the
spectra at different flux levels, defined by the lightcurve of
Fig.~\ref{fig.lightcurves}: A and B are the first and second half of the Jan
2002 observation, respectively, and C and D are the first and second interval of
the Jul 2003 observation. Expressed in units of $10^{-14}$~W\,m$^{-2}$, the
0.5--2.0~keV fluxes of these spectra are 0.44, 1.11, 3.41 and 0.66 for A, B, C
and D, respectively.

In Fig.~\ref{fig.lightcurves} we also show the hardness ratio, defined as the
ratio of the count rates in the 2-8~\AA\ and 8-32~\AA\ band passes. The hardness
ratio of the Jan 2002 observation, $HR=0.215(7)$ is higher than that of the Jul
2003 observation, $HR=0.160(3)$, where the number between brackets is the
1-sigma error in the last digit.  The hardness variation between different flux
levels is small, with $HR=0.239(14), 0.203(8), 0.161(3), 0.134(21)$ for
intervals A, B, C and D, respectively.

\subsection{Continuum}

In both observations the continuum cannot be fitted with only a simple power
law. The soft excess can be modeled by adding a black body modified by coherent
Compton scattering \citep{kaastra2}. The parameters of the power law component
of the Jan 2002 and Jul 2003 observations reflect the results found from the
lightcurve and hardness ratio, which show that for the Jan 2002 observation the
normalisation is three times smaller than for the Jul 2003 observation, while
the slope is flatter, i.e.\ the spectrum is harder in Jan 2002 (see
Table\,\ref{table01}).

In contrast the normalisation of the modified black body is the same within the
error in both observations. The black body temperature is higher in the Jul
2003 observation ($kT=0.143$ keV, vs.\ 0.102 keV in the Jan 2002 observation).

While looking at the different flux levels within each observation, we  find
that the modified black body component does not change between intervals A and
B, whereas the powerlaw is harder at the lower flux of A than at the higher flux
of B. In contrast, the change in flux level between C and D is reflected in the
normalisations of the power law, and of the modified black body, whereas the
slope of the power law is the same in C and D.

\subsection{The ionised absorber}

Thanks to the excellent spectral resolution of the LETG instrument the presence
of a warm absorber in the spectrum is not controversial. We fitted the warm
absorber with model XABS in SPEX. In this model, the ionic column densities are
not independent quantities, but are linked via the ionisation parameter
$\xi=L/nr^2$, where $L$ is the source luminosity, $n$ the hydrogen density and
$r$ is the distance from the ionising source. In addition to $\xi$, the fit
parameters are the hydrogen column density of the absorber $N_H$, the outflow
velocity $v_{out}$ and the turbulent velocity of the medium $v_{turb}$. The
advantage of the XABS model is that all relevant ions are taken into account
including also the ones with the weakest absorption features. 

\subsubsection{The warm absorber in the January 2002 observation\label{warm
January 2002}}

During the January 2002 observation the source is in a low flux state and
consequently the S/N ratio is lower. This has a negative effect on the detection
of the fainter absorption lines which cannot be detected. Only a few strong
absorption lines can be used for the determination of the warm absorber
properties.

With XABS we find that two warm absorber components are needed. The component
with the higher velocity ($-$4450 km\,s$^{-1}$) has a higher ionisation
parameter and a higher column density than the component with the lower outflow
velocity ($-$620 km\,s$^{-1}$; see Table\,\ref{table01}). The absorption system
with the higher outflow velocity improves the fit by $\Delta \chi^{2}=76$ for 4
additional degrees of freedom. This is the fastest outflow velocity yet found in
a Seyfert 1 galaxy. 

We fit the spectra A and B starting with the warm absorber components found in
the total spectrum of Jan 2002.  The separate spectra for A and B then are found
to have within the error bars the same parameters for the warm absorber
components as the total Jan 2002 spectrum. Due to the large uncertainty of the
parameters for spectrum A, we are not very sensitive to detect changes; thus
while the factor of 3 luminosity difference between A and B could in principle
give rise to a similar change of the warm absorber properties, we cannot detect
such a change.

\subsubsection{The warm absorber in the July 2003 observation}

To fit the Jul 2003 observation we need three absorption systems. Two of these
are virtually identical to the components fitted in the Jan 2002 observation;
the third component has a lower outflow velocity ($-$420 km/s), and also a lower
ionisation parameter and lower column density than the other two (see
Table\,\ref{table01}). 

In spectrum D we detected only one absorption system, at an outflow velocity of
about $-$790 km\,s$^{-1}$. The absorption column is higher for spectrum D than
the absorption column for the warm absorber component with the same velocity
detected in interval C and in the total Jul 2003 observation; and the ionisation
parameter is lower. 

\subsection{Broad emission lines}

Besides absorption lines also strong emission features are present in every
spectrum. We first describe the broad emission lines.

\begin{figure}[!htb]
\begin{center}
\includegraphics[angle=-90, width=0.45\textwidth]{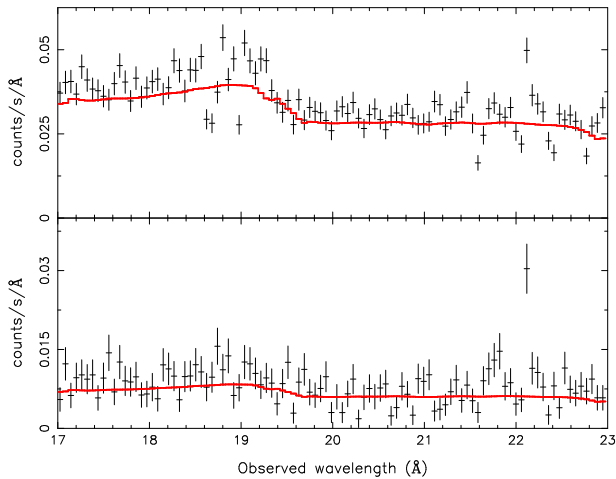}
\caption{Data and continuum model (power law and modified black body
spectrum) in the \mbox{\ion{O}{viii} Ly$\alpha$}
region for the C part (upper panel) and the D part (lower panel). Here the
absorption and emission line features are not yet taken into account.
Note the broad excess near 19 and 22~\AA.} 
\label{broad_line}
\end{center}
\end{figure}

The presence of broad emission features is illustrated in Fig.~\ref{broad_line}
which shows excess flux above the power law and modified black body continuum
for the Jul 2003 observation. Clear excess emission in a band with a width of at
least an \AA{} is present near 19~\AA{} and 22~\AA. This is the region of the
1s--2p transitions in \ion{O}{viii} and \ion{O}{vii}, respectively. In addition,
there is excess emission near 34~\AA{}, the region of the 1s--2p transition in
\ion{C}{vi}. The Jan 2002 observation also shows a significant excess but only
near the \ion{O}{viii} transition.

Close to the black hole the formation of an emission line is influenced by the
strong gravity field and therefore the final emission line has a relativistic
line profile. We modeled this relativistic profile as a narrow emission line 
convolved with the relativistic disk line profile of \citet{laor}. The inner
($r_1$) and outer ($r_2$) disk radius were fixed in our models to the default
values ($r_1=1.234GM/c^2$ and $r_2=400GM/c^2$). The disk inclination $i$ and the
disk emissivity slope $q$ (emissivity proportional to $R^{-q}$ with $R$ the
distance to the black hole) were allowed to vary. In the case when the emission
line is created further away from the black hole this profile shows no or only a
small deviation from the classical gaussian profile.

In order to model the broad emission lines and to search for possible time
variability in particular parts of the observation we proceeded analogously as
in the case of the absorption features. First we determined the best model for
the emission lines for the full observations and after establishing that
emission model we used it as the first approximation for the emission model in
particular parts. 

The parameters that we found for the broad lines are shown in
Table~\ref{table01}. 


\subsection{Narrow emission features\label{Narrow features}}

In both full observations we detect a narrow \ion{O}{vii} forbidden line. The
\ion{O}{vii} forbidden line at 22.10 \AA{} is also visible in each individual
spectrum and has the same strength in spectra A, B and C, but it is stronger in
spectrum D. The line shows only a weak blueshift: the average outflow velocity
is $-110\pm 50$~km\,s$^{-1}$.

The main difference between spectrum D and the other spectra is the presence of
strong Radiative Recombination Continua (RRCs) from \ion{C}{vi} (25.30 \AA) and
\ion{C}{v} (31.62 \AA) in the spectrum, see Fig.~\ref{spectrum_ABCD} and
Table~\ref{table01}.  The temperature of the recombining photoionised gas is
about $5$~eV. In addition the RRCs show a redshift with respect to the rest
frame of the galaxy of $\sim 1000$~km\,s$^{-1}$. We do not find significant
detections for RRCs of \ion{N}{vi} (22.46 \AA), \ion{N}{vii} (18.59 \AA),
\ion{O}{vii} (16.77 \AA) and \ion{O}{viii} (14.23 \AA). In all other spectra we
do not detect significant RRCs, so the numbers given in Table~\ref{table01}
should be interpreted as upper limits.

\section{Discussion}

\subsection{The sudden decline of the source flux in 2003}

In this section we focus upon the sudden decline of NGC~4051 in the July 2003
observation, at the boundary of part C and part D (Fig.~\ref{fig.lightcurves}).
Within 3000~s, the flux dropped by a factor of $\sim 5$ and remained low for
more than 20\,000~s. Large flux variations are common in this object. In fact,
around 29~ks after the start of the Jul 2003 observation, also a drop in
intensity of a factor of $5$ occurred, but  this was after a rise by a factor of
2.5 during the previous 2000~s, and as Fig.~\ref{fig.lightcurves} shows, the
recovery to higher flux level was much faster in that case.

The uniqueness of the transition between part C and D is the rather abrupt
change preceded and followed by long periods of relative quiescence. This makes
it possible for us to see the response of the continuum spectrum, the broad line
emission and the warm absorber to this sudden flux drop. 

The main change between C/D can be characterised as follows (see also
Table~\ref{table01}):
\begin{enumerate}
\item The modified black body component becomes somewhat hotter but with three
times less bolometric flux. If we assume that this component represents the
direct emission from the accretion disk, this can be explained if the inner part
of the disk has disappeared and therefore does not radiate anymore. The only
modest change in temperature then may imply that the range of disk temperatures
in the disappeared part of the disk is not too different from that of the
remaining part.
\item The drop in power law flux is stronger, a factor of $\sim$6, while the
photon index does not change. The disappearance of the inner part of the disk
leads to less seed photons for the assumed inverse Compton scattering corona,
not only because of the lower disk flux ($\sim$3 time less disk flux) but also
due to the relatively lower fraction of disk photons from the outer, remaining
parts of the disk that are captured by a hot corona that is close to the black
hole. The fact that the photon index hardly changes implies probably that
apparently the optical depth and temperature of the corona remains the same
\item The relativistic \ion{O}{viii} line that can be seen in spectrum C is not
visible in spectrum D. This also agrees with the scenario of sudden
disappearance of the inner accretion disk.
\end{enumerate}

We therefore suggest that the sudden decline between part C and part D is caused
by the disappearance of the inner accretion disk. Such phase transitions are not
uncommon in other galactic and extragalactic accretion disk systems. Whether the
disappearing part of the disk is blown away through a strong wind, through jets
or is being accreted onto the black hole is unclear. 

The precise fraction of the disk that disappeared is a little hard to estimate
and strongly model dependent; we only measure that about 2/3 of the bolometric
modified black body flux disappears. Around a maximally rotating disk, 2/3 of
the the \emph{emitted} disk flux comes from within $\sim 2GM/c^2$, but due to
general relativistic effects (gravitational redshift, cone of avoidance causing
most photons to be re-captured by the disk or black hole) this constitutes much
less than 2/3 of the \emph{observed} disk flux. On the other hand, for a
non-rotating black hole, with a smaller cone of avoidance, 2/3 of the flux
originates from within $\sim 20GM/c^2$. Given this, we estimate as an order of
magnitude that the disk within $\sim 10GM/c^2$ has disappeared. With a black
hole mass of $M=3\times10^5 M_\odot$ \citep{mchardy}, the light crossing time
for the diameter of the disappeared region is 30~s. We observe that the fading
takes a full 3000~s, hence the average disappearance speed is only
3000~km\,s$^{-1}$. 

\subsection{The ionised absorber}

\subsubsection{The high ionisation component 2}

In our highest quality spectrum, part C of the July 2003 observation, we detect
three absorption systems (Table~\ref{table01}). Two of these are at a relatively
low velocity (components 1 and 2) while component 3 has a high velocity.

The X-ray component 2, at an outflow velocity of -610~km\,s$^{-1}$, has a column
density of $\sim 2\times 10^{25}$~m$^{-2}$.  Apart from several L-shell lines
from iron ions, its dominant imprint on the absorption spectrum are the
Ly$\alpha$ transitions of hydrogenic oxygen, nitrogen and carbon. Column
density, outflow velocity as well as average 2--10~keV flux of spectrum C are
very similar to the values found by \citet{pounds} in their analysis of the RGS
spectrum taken in May 2001. Therefore we identify our component 2 with their
dominant component. However, in our case the ionisation parameter $\xi$ is a
factor of 10 larger than the value obtained by \citet{pounds}. Various processes
may contribute to this. First, while the 2--10~keV flux in both these
observations are the same, our photon index is $\sim 0.4$ larger, leading to a 
much softer spectrum. For spectrum C the total ionising flux may be a factor of
4 higher, leading to a much higher ionisation parameter for gas at the same
distance and density. A further difference is that during our observation the
relative variability during observation C was relatively small
(Fig.~\ref{fig.lightcurves}), while during the RGS observation the flux varied
wildly (see for instance the light curve shown by \citet{mchardy}). Therefore,
the ionisation parameter found by \citet{pounds} might be an average between
high and low states, if the warm absorber responds fast enough to the continuum
variations. An indication for at least some spread of the ionisation parameter
due to variability in the RGS data is that \citet{pounds} also find a high
ionisation component ($\log\xi = 2.7$) at the same outflow velocity. In our more
"quiet" spectrum C, we do not find evidence for such a highly ionised component
at the outflow velocity of $-600$~km\,s$^{-1}$.

\begin{figure}[!htb]
\begin{center}
\includegraphics[angle=-90, width=0.45\textwidth]{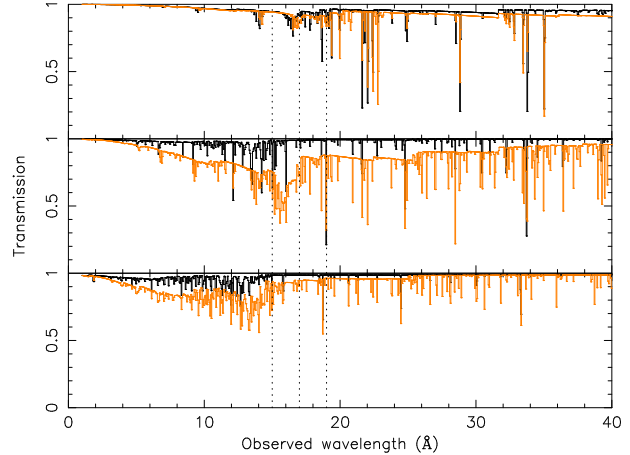}
\caption{Comparison between the model transmission of individual warm absorber
components for the C and D states. From top to bottom: components 1, 2 and 3.
Dark curves: transmission for the high state spectrum C based on the parameters
from Table~\ref{table01}. Light colour curves: for component 2, the transmission
during the low state D; for  components 1 and 3, the transmission for the high
state C but with $\xi$ 5 times lower, therefore showing the expected
transmission if each component would respond very fast to the factor of 5 flux
decrease between C and D. The dotted lines indicate the 15--17 and 17--19~\AA\
bands used in Fig.~\ref{timeresolve}.} 
\label{transmission}
\end{center}
\end{figure}

\begin{figure}[!htb]
\begin{center}
\includegraphics[angle=-90, width=0.45\textwidth]{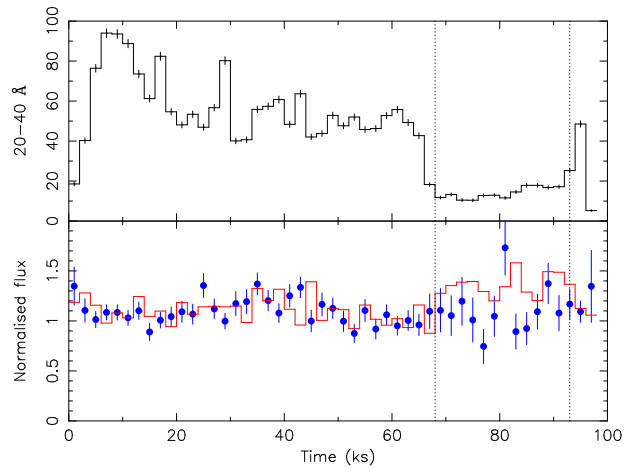}
\caption{Upper panel: light curve in the 20--40~\AA\ band of the July 2003
observation; bin size: 2000~s. Lower panel: fluxes in the 15--17~\AA\ band (data
points with error bars) and 17--19~\AA\ band (histograms; error bars similar to
the other band); these narrow-band fluxes have been normalised to the average
flux per \AA\ in the 20--40~\AA\ band shown in the upper panel. Therefore if
there would be no spectral changes, the normalised 15--17 and 17--19~\AA\ fluxes
should have remained constant.  The dotted lines indicate the boundary of the
low state D.} 
\label{timeresolve}
\end{center}
\end{figure}

Interestingly, when the continuum flux drops by a factor of five in $\sim
3000$~s, defining the boundary between spectrum C and D, the ionisation
parameter of component 2 drops by a similar factor ($8\pm 3$). Because of the
fortunate circumstance that the flux variations during intervals C and D were
relatively small, we therefore conclude that the warm absorber responds rapidly
to the flux decrease. 

How fast does the absorber respond? To that aim, we have plotted the
transmission of component 2 for parts C and D (Fig.~\ref{transmission}, central
panel). The most striking features is a broadband transmission decrease of
10--20~\%. This by itself is not straightforward to measure in spectra with
relatively poor statistics extracted for short time intervals, because a slight
change in continuum parameters could mimic the transmission change. Also, the
signal to noise ratio in individual lines is too low. However,
Fig.~\ref{transmission} shows that there is a sharp decrease in the 15--17~\AA\
band, caused by a deepening of the \ion{O}{vii} edge and the development of the
Fe-M Unresloved Transition Array (UTA) complex between 15--16~\AA. Therefore we
have made light curves in the 15--17~\AA\ band (Fig.~\ref{timeresolve}). A
comparison with the 17--19~\AA\ light curve shows that both bands follow each
other quite well during the high state C, but in the low state D the 15--17~\AA\
flux is on average 25~\% lower than the flux above the \ion{O}{vii} edge
(17--19~\AA\ band), in agreement with Fig.~\ref{transmission}.

It is clear from Fig.~\ref{timeresolve} that the absorber responds definitely
within 10\,000~s to the continuum change; it may be much faster (down to the
1000~s scale or even faster), but the quality of our data does not allow us
to prove that.

We have estimated recombination time scales for arbitrary ions $i$ from the
general equation:

\begin{equation}
\frac{{\mathrm d}N_i}{{\mathrm d}t} =
\Bigl[ -N_i I_i(L) - N_i R_i(T) + N_{i-1} I_{i-1}(L) + N_{i+1} R_{i+1}(T)
\Bigr] n_{\mathrm e}
\label{eqn:ionis}
\end{equation}

where $N_i$ is the density of ion $i$, $n_{\mathrm e}$ the electron density,
$R_i(T)$ the temperature-dependent recombination coefficient from ion $i$ to ion
$i-1$, and $I_i(L)$ is the luminosity-dependent ionisation coefficient from ion
$i$ to ion $i+1$. For our order of magnitude estimate is is sufficient to
neglect here any change in the shape of the ionising spectrum but to reckon only
with the drop in luminosity $L$. We start with the high-state equilibrium
($\log\xi = 2.44$) and calculate the ion concentrations. We then decrease $L$ by
a factor of 5 and evaluate (\ref{eqn:ionis}) and determine the characteristic
time scale $\tau_i$ defined by ${\mathrm d}(\ln N_i)/{\mathrm d}t \equiv
1/\tau_i$. In the high state (C), component 2 has an equilibrium temperature
(for a stationnary state) of 20~eV. In the low state (D), component 2 has an
equilibrium temperature of 5~eV. We therefore determine $\tau_i$ for
temperatures between 5 and 20~eV.  We find that \ion{O}{vii} has a
characteristic time scale $\tau_i$ of $10^4/n_{12}$~s for $T=20$~eV, and
$4\times 10^3/n_{12}$~s for $T=5$~eV, where $n_{12}$ is the hydrogen density in
units of $10^{12}$~m$^{-3}$. For other ions, in particular \ion{Fe}{xiii} --
\ion{Fe}{xvii}, the main contributors to the Fe-M complex between 15--16~\AA\ in
spectrum D, the time scales are 0.2 to 2.5 times the time scale for
\ion{O}{vii}.

We conclude that from our observed upper limit of the recombination time scale
of $t_{\mathrm{rec}}\la 10^4$~s, it follows that the density $n\ga
10^{12}$~m$^{-3}$. Using the definition of the ionisation parameter
($\xi=L/nr^2$), we then obtain an upper limit $r\la 10^{15}$~m for  the distance
of the warm absorber of component 2 to the central source. From its measured
column density of $\sim 2\times 10^{25}$~m$^{-2}$, we find then an upper limit
to the thickness $d$ of the absorber of $d\la 2\times 10^{11}$~m.

The distance upper limit of $10^{15}$~m implies that we can rule out an origin
in the narrow emission line region for component 2. An origin at the inner edge
of a dusty torus ($\ga 3\times 10^{14}$~m, cf. \citet{elvis_escorial}) cannot be
fully ruled out but is unlikely given the fact that our recombination time is
probably faster than $10^4$~s). Therefore,  an origin in the outer parts of an
accretion disk seems more appropriate. The relative thickness of the wind
$d/r\la 2\times 10^{-4}$ is extremely small; this suggests that we have a very
narrow structure, or if the outflow covers a larger region, it must have a small
volume filling factor.

We note that in a recent paper \citep{elvis_escorial} a recombination time of
only 3000~s is proposed based on broad-band fits of time-resolved, low
resolution EPIC spectra. That detection lacks sufficient proof because it is
based upon a correlation study of derived ionisation parameter versus count
rate; at low spectral resolution, however, it is not evident how to disentangle
possible rapidly changing relativistic lines (\citet{ogle} found such lines in
the same XMM-Newton dataset) from a rapidly changing warm absorber. In our case,
we use high-resolution spectra and have a much cleaner transition from a high to
a low state.

\subsubsection{Low ionisation component 1}

At only a slightly smaller outflow velocity ($-200$~km\,s$^{-1}$) compared to
component 2, we find our second, lowly ionised component in spectrum C
(Table~\ref{table01}, component 1).  This component is most easily seen through
the deep 1s--2p absorption lines of \ion{O}{vii}, \ion{O}{vi} and \ion{O}{v}.
The outflow velocity of both components 1 and 2 coincide with the broad troughs
seen in the FUSE \ion{O}{vi} line \citet{kaspi} and in the range of UV
absorption components seen in several lines observed by STIS \citet{collinge}.
Component 1 has five times smaller column density than component 2, and is in
good agreement with the low ionisation tail seen in the column density versus
ionisation plot of \citet{ogle}.

We cannot see component 1 in the low state spectrum D, after the drop of a
factor of 5 in intensity. The five times lower flux level combined with the
shorter exposure time makes detection of any absorption component in spectrum D
harder as compared to spectrum C. Further, even if the gas of this component
responded similarly to the continuum decrease as component 2, it would be hard
to find: Fig.~\ref{transmission} shows that the continuum opacity in both
situations is low for component 1, and also there is no deep Fe-M UTA through
that could be recognised after substantial binning of the data. We therefore
cannot make any statement whether component 1 remained present, or that it got a
lower ionisation.

\subsubsection{High-velocity ionisation component 3}

A high velocity component at an outflow velocity of $-2340$~km\,s$^{-1}$ was
first found by \citet{collinge}; we do not confirm that detection but instead
find a much higher outflow velocity component at $v=-4500$~km\,s$^{-1}$, as
reported fist by \citet{vandermeer}. This component appears to be present in
both our spectra of 2002 and 2003, with almost the same column density and high
ionisation parameter ($\log\xi = 3.2$). In general, this component has a low
opacity. Its predicted response due to the flux decrease between part C/D in the
July 2003 observation is shown in Fig.~\ref{transmission}. If the warm absorber
in this component would be of sufficient low density (we estimate $n\la
10^{14}$~m$^{-2}$), it would respond rapidly to the flux decrease. That would
result in a significant decrease of the transmission of about 20~\% below the
\ion{O}{viii} edge around 14~\AA. We do not see clear signatures of this in our
spectra, but this may be hidden due to the lower opacity than component 2, and
perhaps also a part of the enhanced opacity (if present) could be mimicked in
our spectral fits by a slightly different slope for the power law component in
spectrum D. In the absence of clearer evidence, the distance of component 3
remains unclear.

\subsection{Broad emission lines}

While in the observation of July 2003 we detect with $>3\sigma$ significance
three broad emission features (\ion{O}{viii} Ly$\alpha$, the \ion{O}{vii}
triplet and \ion{C}{vi} Ly$\alpha$) in the observation of January 2002 only one
broad emission feature is present (the \ion{O}{viii} Ly$\alpha$ line). Adding a
broad \ion{C}{vi} Ly$\alpha$ line or the \ion{O}{vii} triplet does not improve
the fit significantly.

There is a distinct difference between the line profile of the broad
\ion{O}{viii} Ly$\alpha$ line in both observations. In the observation of
January 2002, the line is extremely broad (FWHM 21~\AA, with a sharp blue
cut-off near 17~\AA\ and an extended red tail. This line profile is caused by
the steep emissivity law ($q=5.2$) combined with the small inner radius of the
disk (corresponding to a rapidly rotating Kerr black hole). The emissivity slope
$q=5.2$, the inner disk radius $r_i$ and the disk inclination $i=50\deg$ are in
excellent agreement with the values given by \citet{ogle}. These authors
observed NGC~4051 with the RGS instrument of XMM--Newton in May 2001, half a
year before our observation. There is only a small difference in equivalent
width: we find 4.7~\AA\ while Ogle et al. find 2.6~\AA. We conclude that in
general the parameters of the relativistic line as found by us in the Jan 2002
spectrum are rather similar to the parameters found by Ogle et al. 

However, Ogle et al. modeled the soft excess in NGC~4051 with the help of
relativistic emission from the entire \ion{O}{viii} emission lines series and
RRCs.  We tried to model the soft excess in the RGS (2-37\AA) spectral band with
the same model used by \citet{ogle} (i.e., ignoring the modified black body
component). This trial was not successfully neither for the January 2002
observation nor for the July 2003 observation (see Table \ref{table03}). We
conclude that in both LETGS observations the soft excess requires also the
presence of the modified black body emission, rather than the higher Lyman
transitions of \ion{O}{viii}.

During the observation of July 2003 the profile seems to be produced under
different conditions. We tested the possibility that also in this observation
the \ion{O}{viii} Ly$\alpha$ profile is modified by relativistic effects. The
inclusion of a relativistic broadened profile in the model is not statistically
more significant than a simple gaussian profile. However, a gaussian model
predicts a velocity broadening of the line of the order of
$30\,000$~km\,s$^{-1}$ and this velocity broadening may hide some relativistic
smearing. The best fit for a relativistic profile is achieved with an emissivity
slope $q=2$, with a FWHM of 2.5~\AA. Thus the line profile is an order of
magnitude narrower than in the observation of January 2002. A similar behaviour
of the emissivity slope $q$, as a function of the central source flux, was found
also for another narrow line Seyfert~1 galaxy (\mbox{MCG~$-$6-30-15},
\citet{iwasawa}), but in that case for the Fe-K line. Considering the same
scenario as for \mbox{MCG~$-$6-30-15}, the red tail during January 2002
observation of NGC~4051 suggests that during that low state the line should be
produced very close to the black hole where gravitational effects are
significant. The blue horn is gravitationally shifted into the red wing and
almost all of the line emission goes to the red wing. Conversely, during a high
state most of the emission comes from a larger distance. Thus the blue peak
comes mostly from the blue side of the disk and the red wing is depressed
(\citep{iwasawa}).

The spectrum in July 2003 is dominated by the high state C. Surprisingly, after
the sudden drop in intensity that occurs at the end of spectrum C, an extremely
broad line as seen during the lower flux states A and B does not re-appear.
Instead, there remains only a weak, relatively narrow broad emission line (FWHM
0.69~\AA\ or 11\,000~km\,s$^{-1}$), which can be attributed to the outer
accretion disk or the broad line region.

The fact that the \ion{O}{viii} Ly$\alpha$ line may be explained with a
relativistic profile while the \ion{O}{vii} triplet has a gaussian profile does
not to need be contradictory. Most of the \mbox{\ion{O}{viii} Ly$\alpha$} line
can be formed by the gas which is closer to the black hole and the gas
producing the \ion{O}{vii} line may be placed further away. However an important
caveat is that \ion{O}{vii} triplet shape is more difficult to determine as it
is a blend. 

\begin{table}
\begin{center}
\caption{\label{table03}Comparison of the soft excess models for 
the January 2002 and July 2003 observation.}
 \begin{tabular}[t]{llcc}
   \hline
   \hline
Observation   &model		      &$\chi^2$       &dof\\
\hline
January 2002  &relativistic lines model$^a$ &$698$	      &$561$\\
	      &our best model$^b$           &$684$	      &$565$\\
\hline
July 2003     &relativistic lines model$^a$ &$1640$	      &$1148$\\
	      &our best  model$^b$          &$1326$	      &$1152$\\
\hline
\end{tabular}
\end{center}
\noindent
$^a$ the relativistic lines model consists of a power law 
plus the entire series of \ion{O}{viii} lines
broadened by the relativistic profile as in \citet{ogle}\\
$^b$ our best model consists of a power law, modified black body and
a single relativistic \ion{O}{viii} Ly$\alpha$ line.
\end{table}

Relativistic line profiles in the soft X--ray band have been discovered first by
\citet{branduardi} in some other narrow lines Seyfert 1 galaxies
(\mbox{MCG~$-$6-30-15} and Mrk~766). In these sources, the disk line parameters
derived from these lines and the Fe-K line are in good agreement.  The
XMM-Newton observation of NGC~4051 that showed a strong relativistic
\ion{O}{viii} line \citep{ogle} did not show a significant relativistic Fe-K
line \citep{pounds}. Unfortunately the LETGS is not sensitive enough in the the
Fe-K band and thus we cannot directly compare the disk geometry  as derived from
iron and oxygen lines. For the physical implications of the relativistic lines
in NGC~4051 we refer to \citet{ogle}.

\subsection{Narrow features}

During our observations we detect various narrow emission features in the
spectra. The forbidden \ion{O}{vii} line and RRCs of \ion{C}{v} and \ion{C}{vi}
were the most dominant.

The \ion{O}{vii} forbidden line is narrow and does not respond to the flux
change (Table \ref{table01}). This suggests that this line is produced by gas
away from the central source, for example in the narrow line region.

There is no evidence for any variation of the centroid of the \ion{O}{vii}
forbidden line. Combining all observations, the centroid of $22.093\pm
0.004$~\AA\ as measured in the restframe of NGC~4051 corresponds to a net
outflow velocity of $-110\pm 50$~km\,s$^{-1}$.

The strongest RRCs are detected during a short time interval in the July 2003
observation (Fig. \ref{spectrum_ABCD}). The shape and centroid of individual
RRCs is not only affected by the temperature of the recombining gas and the
emission measure of the relevant ion, but also by the motion (velocity
broadening) of the recombining gas. In a previous part of our paper (section
\ref{Narrow features}) we determined the temperature and emission measure
assuming zero velocity broadening. How can the velocity broadening affect the
determination of the temperature and the emission measure? We have investigated
this by including in our RRC spectral model a velocity broadening component. 
On the basis of the goodness of fit ($\chi^2$) we exclude that the recombining
gas is moving faster than $3\,000\,\mathrm{km\,s}^{-1}$. The temperature and
emission measure are within the error bars the same for models with a velocity
broadening of  $3\,000\,\mathrm{km\,s}^{-1}$ and for models without velocity
broadening.

\begin{figure}[!htb]
\begin{center}
\includegraphics[angle=-90, width=0.45\textwidth]{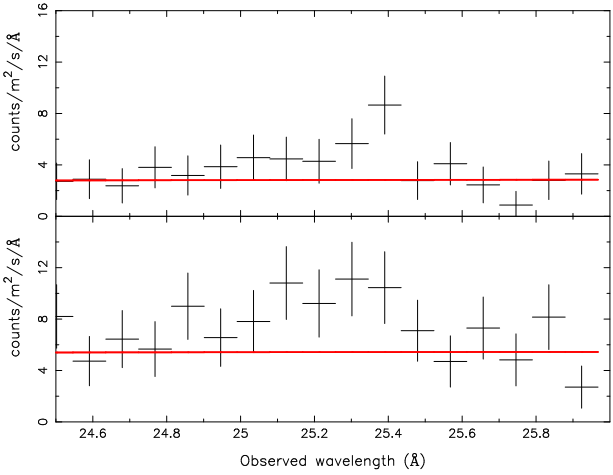}
\caption{Data and continuum model in the \ion{C}{vi} RRC region for
the first half (upper panel) and the second half (lower panel) of
interval D (see Fig.  \ref{fig.lightcurves}) of our observation of
NGC~4051.}
\label{lowrrc}
\end{center}
\end{figure}

If we interpret the velocity broadening as a typical Keplerian velocity at the
location where the RRC is formed, we can derive some properties of the
recombining gas. For a black hole mass of $M=3\times10^5 M_\odot$
\citep{mchardy} and the maximum velocity broadening of $\Delta
v=3000\,\mathrm{km\,s}^{-1}$, we typically get a distance of the recombining gas
of $R\sim 4\times 10^{12}$ m or $\sim 4500$ Schwarzschild radii.

From the fitted temperature of the RRC during part D ($4.9\pm 1.8$~eV) we derive
an ionisation parameter $\log \xi = 1.6\pm 0.3$. With the ionising luminosity of
the source ($L=2.8\times 10^{35}$ W), our distance derived above and the
definition of $\xi$ we obtain a typical density of $4\times 10^{17}$~m$^{-3}$.
It is now also possible to estimate the volume of the RRC emitting source. Using
the observed emission measure of the \ion{C}{vi} RRC of $3.2\times
10^{66}$~m$^{-3}$, a cosmic carbon abundance and a fraction of 50~\% of the
Carbon in the form of \ion{C}{vii} (appropriate for $\log\xi = 1.6$), we find an
emitting volume of $V=1.2\times 10^{35}$~m$^{3}$. Equating this to $V=\Omega
R^2\Delta R$ with $\Delta R$ the characteristic thickness of the emitter in the
radial direction and $\Omega$ the solid angle sustained by the emitter, we
find   
\begin{equation} 
\Omega \frac{\Delta R}{R} = 0.002 
\end{equation}

Another constraint follows from the column density $N_{\mathrm H}$ through the
emitting region. This can be written as $N_{\mathrm H} = n\Delta R = n R (0.002
/ \Omega)$ or substituting more numbers $N_{\mathrm H} = 2.5\times 10^{26}
(4\pi/\Omega)$~m$^{-2}$. Using that estimate, we can rule out an origin of the
RRC emission in the warm absorber: the ionisation parameter $\log\xi\simeq 1.6$
would suggest an association with warm absorber component 2, but the column
density of that component is at least 10 times smaller. Instead, the high
(column) density may suggest an origin at the thin, ionised skin of the
accretion disk at a few thousand Schwarzschild radii from the black hole. 

We conclude with some consistency checks. First, the light crossing time through
the emitting region is of order $10^4$~s, consistent with the duration of phase
D. Note that we see the RRC during the full interval D (Fig.~\ref{lowrrc}).
Further, for $\log \xi \simeq 1.6$ we expect only a small fraction of gas
recombining into \ion{O}{viii}, but instead we expect a strong \ion{O}{vii} RRC
at 16.77 \AA, but this feature is not clearly observed. However in this spectral
region absorption from ions such as \ion{Fe}{ix} may mask the presence of
\ion{O}{vii} RRC emission.

\section{Conclusions}

In this study we have investigated the spectral properties of NGC~4051, 
observed by {\it Chandra}-LETGS on two occasions for a total of 180\,ks.

The time averaged spectrum of both the Jan 2002 and the Jul 2003 can be fitted
by a modifyed black body and a power law, absorbed by ionised gas. 

Two of the ionisation components (2 and 3) are well visibible both in Jan 2002
and in the high state (part C) of Jul 2003. These gas components are consistent
with being stable over a long time scale ($\sim19$\,months) both in ionisation
level and outflow velocity. 

In particular, we report the detection of the highest outflow velocity-gas ever
observed ($v\sim-4800$\,km s$^{-1}$), for component 3.  The lower ionisation
phase of the gas (component 1) is well detected only in the state C. This lack
of detection is most probably due to the lower statistics affecting the other
time segments.

In the last part of the Jul 2003 observation, (part D), only the second
component is detected, but it shows a significant variation in the ionisation
parameter which linearly responded to the continuum flux variation on a time
scale $>$3000\,s. From this we estimated a lower limit for the gas density
($n\ga 10^{12}$\,m$^{-3}$) and as a consequence a distance for the absorber
$r\la 10^{15}$\,m and a thickness of the gas layer $d\ga 2\times 10^{11}$\,m.

\noindent
The emission spectrum is rich in narrow and broad emission lines:

The narrow lines (e.g. the \ion{O}{vii} forbidden line) are not variable in time
and therefore consistent with being produced in regions very distant from the
black hole (e.g. the NLR).

The RRCs of \ion{C}{v} and \ion{C}{vi} showed instead a very rapid variability,
significantly arising above the continuum after the flux drop in spectrum D. The
estimated distance ($r\sim4\times10^{12}$\,m) of the recombining gas shows that
these RRCs are produced close to (or in) the accretion disc.  

We also find that the column density of the gas producing the RRCs is
incompatible with that found for the warm absorber, suggesting a different
location for the absorbing and emitting media. 

Broad lines of \ion{O}{viii}, \ion{O}{vii} and \ion{C}{vi} are also evident in
the spectrum. The \ion{O}{vii} and \ion{C}{vi} lines show a simple Gaussian
profile with a FWHM ranging from 0.3 to 1.5 \AA, suggesting an origin in the
BLR. 

On the contrary, the \ion{O}{viii} Ly$\alpha$ line is best modeled by a
relativistically broadened profile. The tentatively detected variability  on a
months time scale is according to expectations for a line produced this close to
a black hole. In accordance with previous studies, we find that the observed
anticorrelation between the emissivity slope and the central source flux can be
explained in terms of a relativistic accretion disc around a spinning black
hole.  

\begin{acknowledgements}
SRON
is supported financially by NWO, the Netherlands Organization for Scientific
Research.
\end{acknowledgements}

\end{document}